\documentclass{IEEEtran}
\usepackage{cite}
\usepackage{amsmath,amssymb,amsfonts}
\usepackage{siunitx}
\usepackage{subcaption}
\usepackage{graphicx}
\usepackage{overpic}
\usepackage[justification=centering]{caption} 
\usepackage{textcomp,nicefrac}
\usepackage[switch]{lineno}
\usepackage{placeins}
\def\BibTeX{{\rm B\kern-.05em{\sc i\kern-.025em b}\kern-.08em
T\kern-.1667em\lower.7ex\hbox{E}\kern-.125emX}}
\begin{document}
\title{Stability of Charge Collection Efficiency and Time Resolution in 4H-SiC PIN Diodes Under X-ray Irradiation}
\author{Jiaqi Zhou$^{1,2}$, Sen Zhao$^{2}$, Xiyuan Zhang$^{2,3,\ast}$, Suyu Xiao$^{4}$, Chenxi Fu$^{2}$, Congcong Wang$^{2}$, Yanpeng Li$^{1}$, Weimin Song$^{1,\ast}$, and Xin Shi$^{2,3}$
\thanks{$\ast$ Corresponding authors: Xiyuan Zhang (e-mail: zhangxiyuan@ihep.ac.cn) and Weimin Song (e-mail: weiminsong@jlu.edu.cn).}
\thanks{This work is supported by the National Natural Science Foundation of China (Nos.~12205321, 12375184, 12305207, and 12405219), National Key Research and Development Program of China under Grant No. 2023YFA1605902 from the Ministry of Science and Technology, China Postdoctoral Science Foundation (2022M710085), Natural Science Foundation of Shandong Province Youth Fund (ZR2022QA098) under CERN RD50-2023-11 Collaboration framework.}
\thanks{$^{1}$Jiaqi Zhou, Yanpeng Li, and Weimin Song are with Jilin University, Changchun 130015, China.}
\thanks{$^{2}$Sen Zhao, Xiyuan Zhang, Chenxi Fu, Congcong Wang, and Xin Shi are with the Institute of High Energy Physics, Chinese Academy of Sciences, Beijing 100049, China.}
\thanks{$^{3}$Xiyuan Zhang and Xin Shi are also with the State Key Laboratory of Particle Detection and Electronics, Beijing 100049, China.}
\thanks{$^{4}$Suyu Xiao is with Shandong Institute of Advanced Technology, Jinan, China.}}
\maketitle

\begin{abstract}
This study evaluates the radiation tolerance of a 4H-SiC PIN detector under X-ray irradiation up to \SI{2}{MGy} (Si) at \SI{160}{keV}. The detector features a fully epitaxial vertical PIN structure with mesa terminations and field plates. Comprehensive pre- and post-irradiation characterization includes I-V/C-V measurements, charge collection efficiency (CCE) and timing resolution tests using $\beta$-particles ($^{90}$Sr). After \SI{2}{MGy} irradiation, the reverse leakage current remains at an ultralow level of $\sim 10^{-11}$ \si{A/cm^2} at \SI{-300}{V} with negligible degradation. C-V characteristics are basically consistent, with full depletion at \SI{~130}{V}. CCE for $\beta$-particles decreases by less than 5\%. The detector maintains good timing resolution: \SI{21}{ps} before and \SI{31}{ps} after irradiation, with jitter increasing moderately. These results demonstrate stable performance under extreme X-ray exposure, highlighting the detector's potential for radiation-hard applications in high-energy physics, space missions, and nuclear reactor monitoring.
\end{abstract}


\section{INTRODUCTION}
\label{sec:introduction}

\IEEEPARstart{T}{he} relentless advancement of modern technology, particularly in nuclear energy, space exploration, and high-energy physics, has created an urgent demand for radiation-tolerant X-ray sensors capable of stable and reliable operation in increasingly harsh environments \cite{SELLIN2006479, Karmakar2021}. Unlike active detection systems that emit radiation, the passive sensors discussed herein are designed to be immersed in radiation fields, continuously monitoring and quantifying X-ray intensity or cumulative dose. Their performance is critical for ensuring safety, enabling scientific discovery, and prolonging mission lifetimes. Traditional semiconductor sensors, such as those based on silicon, are highly susceptible to total ionizing dose (TID) effects and displacement damage (DD) when exposed to high-dose radiation. These damages manifest as significant increases in leakage current, degradation of charge collection efficiency (CCE), and eventual device failure, severely limiting their applicability in extreme environments \cite{Moll2018, Ren2024}.

The need for robust, passive X-ray sensors is particularly acute in environments characterized by intense and sustained radiation fields.In nuclear facilities, including power reactors, fuel cycle plants, and waste storage sites, continuous monitoring of X/gamma radiation is paramount for personnel safety and structural integrity assessment. Sensors deployed here must operate stably for decades under high radiation fluxes, providing accurate measurements without being degraded by cumulative effects \cite{Karmakar2021}. Similarly,in high-energy physics (HEP) experiments, detectors at particle colliders and accelerators are exposed to extreme radiation doses while required to precisely timestamp and measure the energy of numerous particles, including X-rays, from collision events. These detectors demand exceptional radiation hardness, high signal-to-noise ratio, and fast response times to function effectively over years of operation \cite{SELLIN2006479, CEPCStudyGroup:2018rmc, DeNapoli2022}.

Beyond these extreme scenarios, radiation-tolerant X-ray sensors are vital in other key areas. Inspace exploration, satellites and probes are subjected to a constant barrage of cosmic rays and solar X-rays, necessitating sensors for navigation, scientific instrumentation, and monitoring of electronic systems against single-event effects (SEE) and TID \cite{WOS:000803113800029, Benoit2016}. Inmedical applications, such as radiotherapy and advanced imaging, precise real-time dosimetry is crucial for maximizing tumor control while minimizing damage to healthy tissue. The growing emphasis on ultra-low-dose X-ray techniques for long-term health monitoring and pediatric imaging further underscores the need for highly sensitive and stable detectors \cite{Li2021, UCSCboard}. Finally, in industrialnondestructive testing (NDT) and security screening, high-intensity X-ray sources used to inspect dense or complex structures require detectors that maintain performance despite significant radiation exposure \cite{Yang2025}.

Addressing these challenges has spurred extensive research into novel materials and device architectures. Wide-bandgap (WBG) semiconductors, such as silicon carbide (4H-SiC), gallium nitride (GaN), and gallium oxide (Ga$_2$O$_3$), have emerged as leading candidates for next-generation radiation-hard detectors \cite{DeNapoli2022, Karmakar2021, Zhang2024}. Their high displacement threshold energy, strong atomic bonding, low intrinsic carrier concentration, and high critical electric field confer inherent resistance to radiation-induced lattice damage and enable stable operation at high temperatures \cite{Harley-Trochimczyk_2017, ion-implant}. While other material systems, such as halide perovskites, show great promise for low-dose and self-powered detection due to their unique optoelectronic properties \cite{Yang2025, Benoit2016}, 4H-SiC stands out for its mature epitaxial growth technology and proven performance in high-field and high-radiation environments.

In this study, we investigate the radiation hardness of a fully epitaxial 4H-SiC PIN diode subjected to high-dose \SI{160}{keV} X-ray irradiation up to a cumulative dose of \SI{2}{MGy} (Si equivalent). Through comprehensive pre- and post-irradiation characterization—including I-V and C-V measurements, charge collection efficiency tests using $\beta$-particles ($^{90}$Sr), and timing resolution analysis—we systematically evaluate the detector's degradation. Our results demonstrate exceptional stability: the device maintains an ultralow leakage current, exhibits minimal ($<$5\%) CCE loss, and, for the first time, shows that an excellent timing resolution of \SI{31}{ps} can be preserved after such extreme exposure. These findings conclusively highlight the potential of 4H-SiC PIN detectors as robust candidates for a wide spectrum of applications demanding both radiation hardness and fast timing, from nuclear reactor monitoring and space missions to future HEP experiments and advanced medical imaging.

\section{Device Configuration and Processing Technology}
\label{sec:device}

\begin{figure*}[!t]
    \centering
    \begin{minipage}[b]{0.23\textwidth}
        \centering
        \begin{overpic}[width=\textwidth]{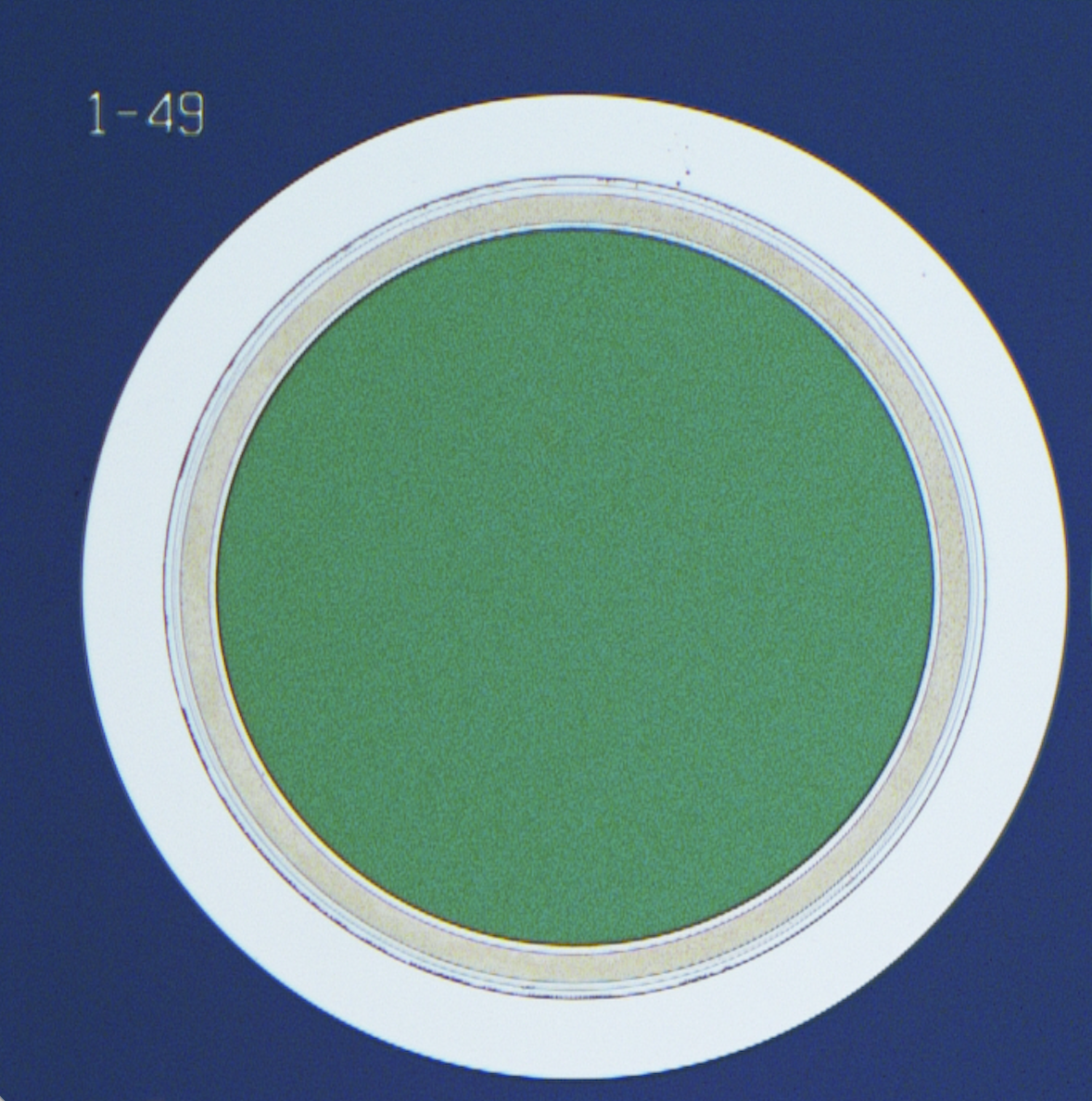}
            \put(-13,98){\textbf{(a)}}
        \end{overpic}
    \end{minipage}%
    \hspace{1.5cm}%
    \begin{minipage}[b]{0.45\textwidth}
        \centering
        \begin{overpic}[width=\textwidth]{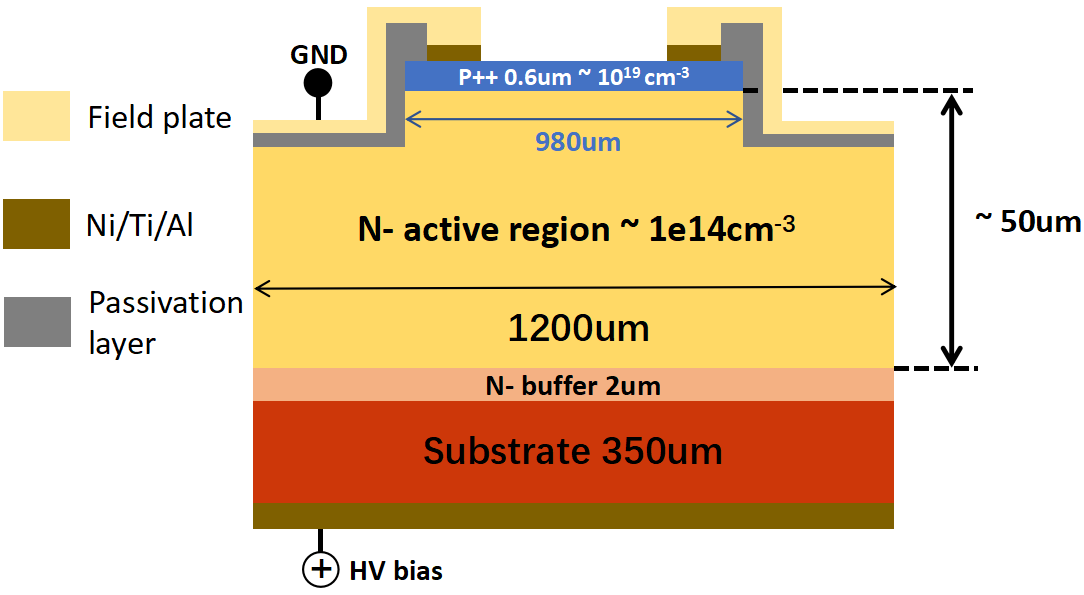}
            \put(0,50){\textbf{(b)}}
        \end{overpic}
    \end{minipage}
    \caption{(a) Schematic diagram of the 4H-SiC PIN detector. (b) Detailed cross-sectional structure.}
    \label{fig1}
\end{figure*}

This study investigates the degradation mechanisms of 4H-SiC PIN diodes under X-ray irradiation. The devices under test were custom-designed and fabricated at the Institute of High Energy Physics (IHEP), Chinese Academy of Sciences. The schematic cross-section of the device is presented in Fig.~\ref{fig1}(a).

The detector employs a fully epitaxial vertical PIN architecture, which enables precise control of thickness and doping profiles while eliminating lattice damage typically induced by conventional high-temperature ion implantation and annealing processes \cite{ion-implant}. The structure comprises: 
\begin{itemize}
    \item A \SI{50}{\mu m}-thick lightly doped N-type (N$^-$) epitaxial layer with a doping concentration of $1 \times 10^{14}$ \si{cm^{-3}}, which serves as the charge-sensitive active region.
    \item A heavily doped P-type (P$^{++}$) epitaxial anode layer with a thickness of \SI{0.6}{\mu m} and doping concentration $>1 \times 10^{19}$ \si{cm^{-3}}.
    \item An N-type conductive substrate serving as the cathode.
    \item The active region has a diameter of \SI{980}{\mu m} (radius \SI{490}{\mu m}).
\end{itemize}

In 4H-SiC, high-dose ion implantation induces progressive lattice mismatch ($\Delta C/C_0 > 0.5\%$ at $1\times 10^{20}$ \si{cm^{-3}}) and defect formation \cite{10.1063/1.4720435}. The epitaxial approach used in this study maintains good lattice coherence with the SiC substrate, avoiding the lattice mismatch and crystallographic tilt induced by high-dose ion implantation.

To ensure high-voltage stability, the device features a mesa termination structure created by dry etching with a sidewall angle of 45$^\circ$--60$^\circ$ and an etch depth of \SI{1.6}{\mu m}. A SiO$_2$ passivation layer was subsequently deposited via Plasma-Enhanced Chemical Vapor Deposition (PECVD) at \SI{350}{\degreeCelsius}. Ohmic contacts were formed by electron beam evaporation of Ni/Ti/Al (50/15/80 nm) multilayer electrodes on both the P$^{++}$ anode and the C-face of the substrate, followed by rapid thermal annealing (RTA) at \SI{850}{\degreeCelsius} for 5 minutes. This annealing condition has been reported to achieve a very low contact resistivity of \SI{6.25e-5}{\Omega\centi\metre\squared} \cite{TempDependent}, satisfying the requirements for PIN device fabrication. A key design feature is the extension of the anode metal over the passivation layer, forming an integrated field plate to suppress electric field crowding at the mesa edge. The anode is patterned into a ring-shaped electrode.

\subsection{Irradiation and Characterization Setup}

The X-ray irradiation experiments were conducted using a MultiRad160 facility (Multi-Rad 160) at the Institute of High Energy Physics (IHEP), Chinese Academy of Sciences. All irradiations were performed in a temperature-controlled environment maintained at \SI{20}{\degreeCelsius}. During exposure, the devices were subjected to a \SI{160}{keV} X-ray beam at a dose rate of \SI{246}{Gy/min} (Si equivalent), with the beam incident normally on the device surface. Four cumulative dose points were investigated: 0, 0.5, 1, and \SI{2}{MGy}.

The electrical characterization of the devices was conducted on a probe station under ambient conditions. For the current-voltage (I-V) measurements, a Keithley 2470 high-voltage source meter was utilized to apply the DC bias and simultaneously record the leakage current. Capacitance-voltage (C-V) profiling was performed at a fixed frequency of \SI{100}{kHz} using a Keysight E4980A Precision LCR meter, which was equipped with an external bias tee to superimpose the AC test signal onto the DC bias voltage supplied by the source meter. The measurement system exhibited a parasitic capacitance of 5--8 pF, which was subtracted during data processing.

A stepped voltage sweep protocol was employed for both measurements to ensure accuracy and repeatability. The bias voltage was incremented in steps of \SI{2}{V} for I-V and \SI{1}{V} for C-V measurements. At each voltage step, the measurement was repeated five times, and the reported value represents the arithmetic mean of these readings, thereby minimizing random noise and enhancing data reliability.

The charge collection and timing resolution measurements were conducted using a Strontium-90 $\beta$-particle source, which provides beta particles with a maximum energy of \SI{2.28}{MeV}. The experimental setup integrated UCSC single-channel readout boards equipped with transimpedance amplifiers (TIA). These boards feature a bandwidth of \SI{1.6}{GHz}, a transimpedance gain of \SI{470}{\Omega}, and an output noise level of approximately 1-1.5 mV (RMS) \cite{UCSCboard}. Device bias was supplied by a Keithley 2470 source meter, while the operating power for the readout board was provided by a GWinstek GPD-3303S low-voltage unit. Signal acquisition was performed with a Tektronix MSO64 oscilloscope, configured with a bandwidth of \SI{2.5}{GHz}, an input impedance of \SI{50}{\ohm}, and a sampling rate of \SI{10}{GS/s} to ensure sufficient signal-to-noise ratio (SNR) for the measurements.

\section{Results and Discussion}
\label{sec:results}

\begin{figure*}[!t]
    \centering
    \begin{minipage}[b]{0.36\textwidth}
        \centering
        \begin{overpic}[width=\textwidth]{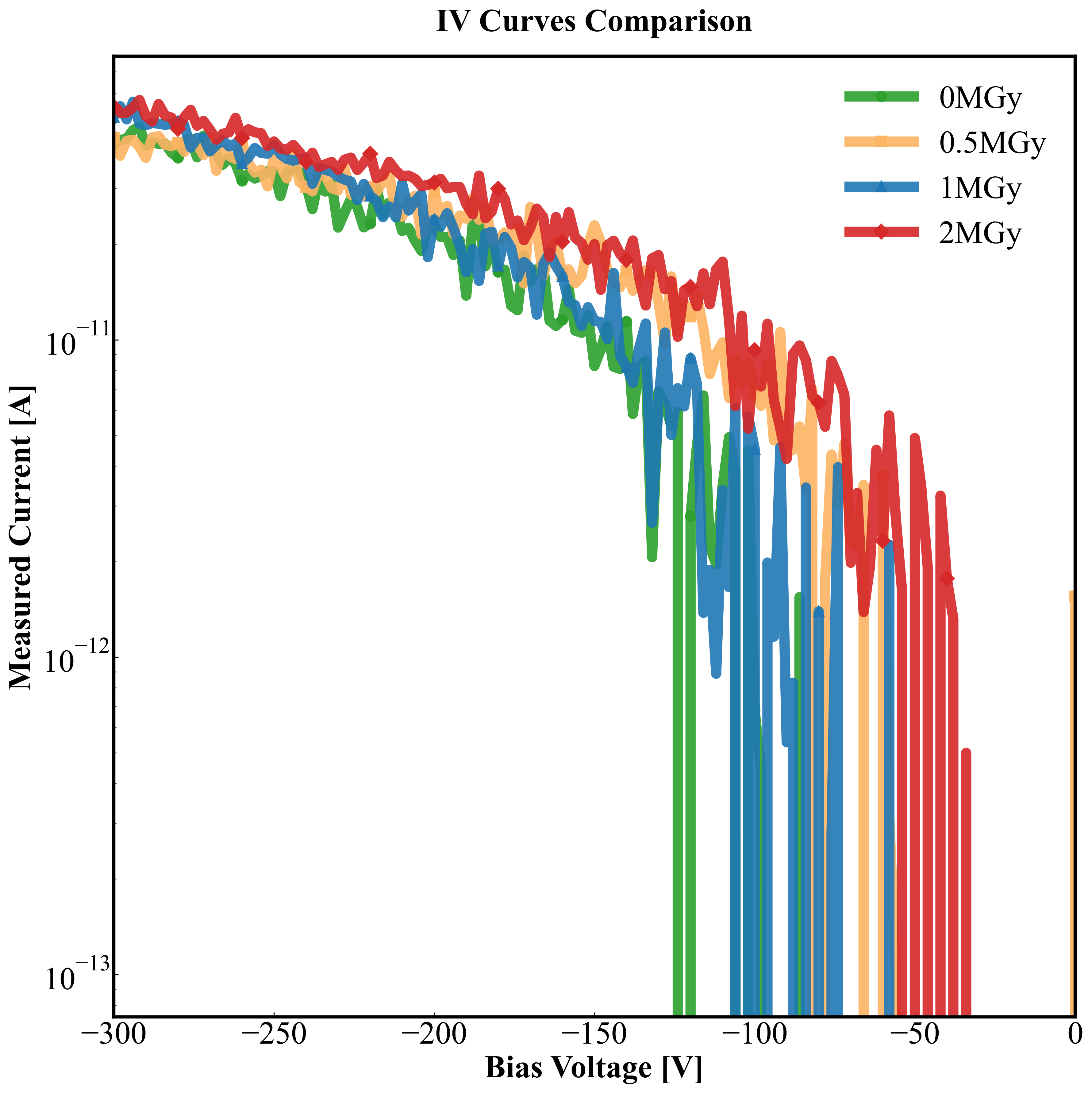}
            \put(0,93){\textbf{(a)}}
        \end{overpic}
    \end{minipage}%
    \hspace{1.5cm}%
    \begin{minipage}[b]{0.36\textwidth}
        \centering
        \begin{overpic}[width=\textwidth]{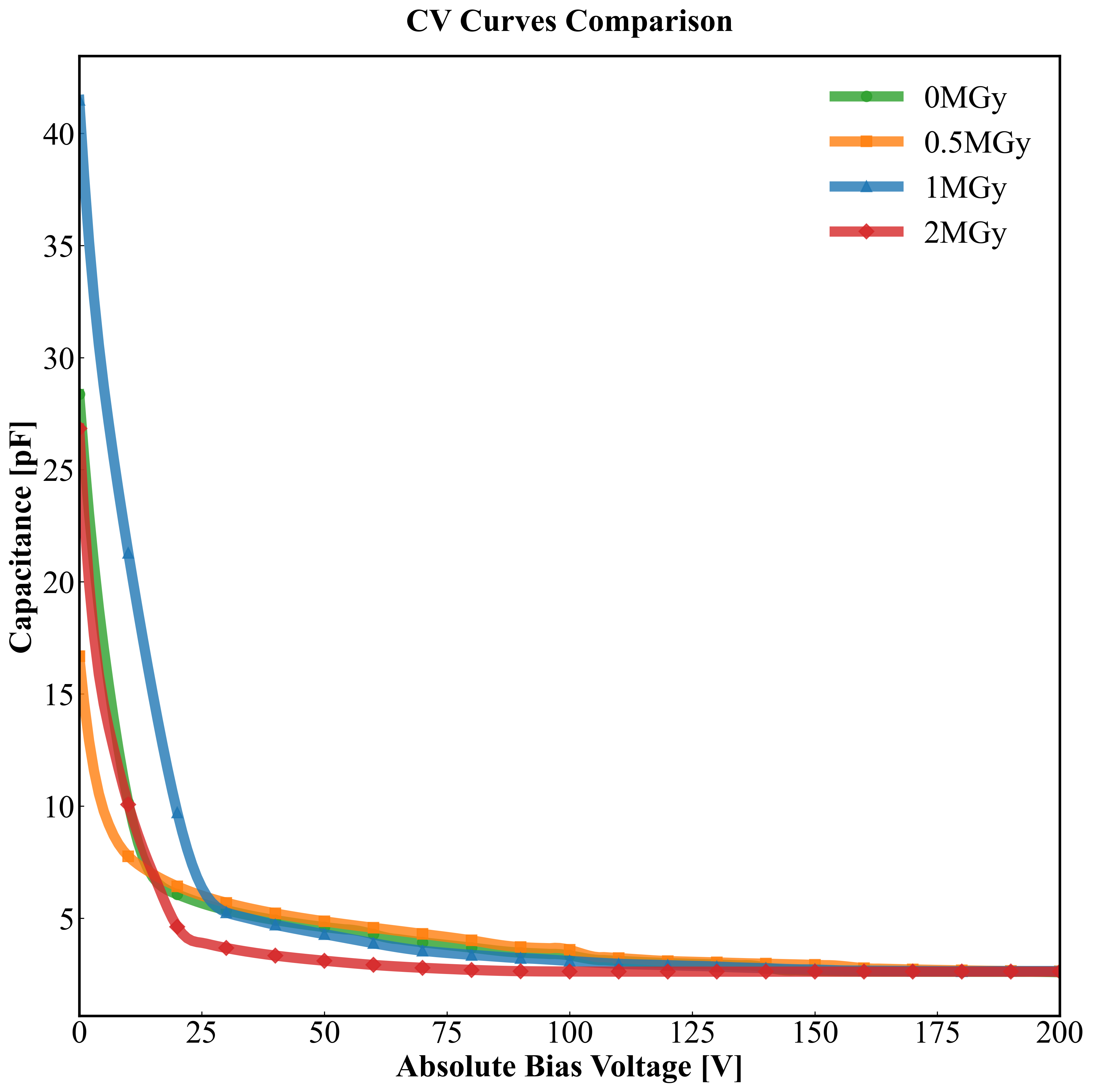}
            \put(-2,93){\textbf{(b)}}
        \end{overpic}
    \end{minipage}
    \caption{(a) Reverse I-V characteristics of 4H-SiC PIN detectors before and after X-ray irradiation at different doses. (b) Reverse C-V characteristics of 4H-SiC PIN detectors before and after X-ray irradiation at different doses.}
    \label{fig2}
\end{figure*}

\subsection{Current-Voltage Characteristics}

To comprehensively evaluate the radiation hardness of the devices, the static current-voltage (I-V) characteristics under different irradiation doses were systematically characterized, as shown in Fig.~\ref{fig2}(a). Under reverse bias, the post-irradiation leakage current curves nearly overlap with the initial ones, showing no current increase attributable to radiation damage. Specifically, even at an extreme reverse bias of \SI{-300}{V}, the leakage current of the device remains effectively suppressed at an ultralow level of $\sim 2 \times 10^{-11}$ A.

This outstanding electrical robustness stands in sharp contrast to typical silicon-based detectors, whose leakage current usually degrades by several orders of magnitude under comparable irradiation conditions due to a dramatic increase in bulk and surface generation currents \cite{Moll2018}. The reverse leakage current can maintain an ultralow level primarily because the wide bandgap of SiC significantly suppresses thermally generated current. More importantly, the interaction of X-rays with matter is dominated by ionization energy loss, resulting in extremely weak displacement damage. The high bonding energy between SiC atoms effectively resists such damage, making it difficult for leakage-inducing defects to form.

\subsection{Capacitance-Voltage Characteristics}

The capacitance-voltage (C-V) characteristics of the PIN detectors before and after X-ray irradiation are shown in Fig.~\ref{fig2}(b). The C-V curves clearly show that all samples—including the unirradiated reference and those subjected to various irradiation doses—enter a plateau region where the capacitance stabilizes once the reverse bias reaches approximately \SI{130}{V}. This feature indicates that the internal electric field has penetrated the entire lightly doped N-drift region (the intrinsic I-layer), signifying the complete depletion of the active area.

Close examination of Fig.~\ref{fig2}(b) reveals minor differences in the low-bias region (0-50 V) among different irradiation doses, leading to slight variations in the onset of the depletion plateau. These minor differences are attributed to inherent process variations between different devices, such as minute deviations in active layer thickness or doping concentration during epitaxial growth, rather than radiation-induced damage. Such variations are within acceptable fabrication tolerances.

Importantly, despite these minor initial differences, the depletion capacitance of all samples converges to the same stable value (approximately \SI{2.6}{pF}) at high bias voltages ($>$150 V). The plateau capacitance after full depletion is primarily determined by the active layer thickness and dielectric constant. This convergence demonstrates that even after exposure to a high X-ray irradiation dose of up to \SI{2}{MGy}, the active layer thickness and basic dielectric properties of the devices remain unaltered. This observation directly confirms, from a capacitance perspective, that the defects potentially introduced by X-ray ionization at this dose level do not markedly alter the effective doping concentration of the drift region, thus not affecting the formation and expansion of the depletion layer.These C-V results confirm the structural integrity of the devices after irradiation, setting the stage for charge collection efficiency measurements discussed next.

\subsection{Charge Collection Efficiency}

\begin{figure*}[!t]
    \centering
    \begin{minipage}[b]{\textwidth}
        \centering
        \begin{overpic}[width=0.77\textwidth]{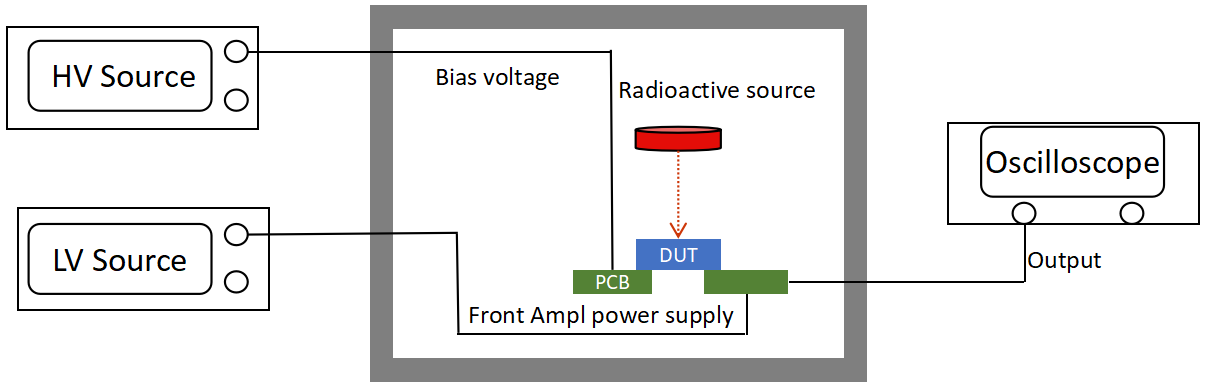}
            \put(-5,29){\textbf{(a)}}
        \end{overpic}
    \end{minipage}\\[2ex]
    
    \begin{minipage}[t]{0.37\textwidth}
        \centering
        \begin{overpic}[width=\textwidth]{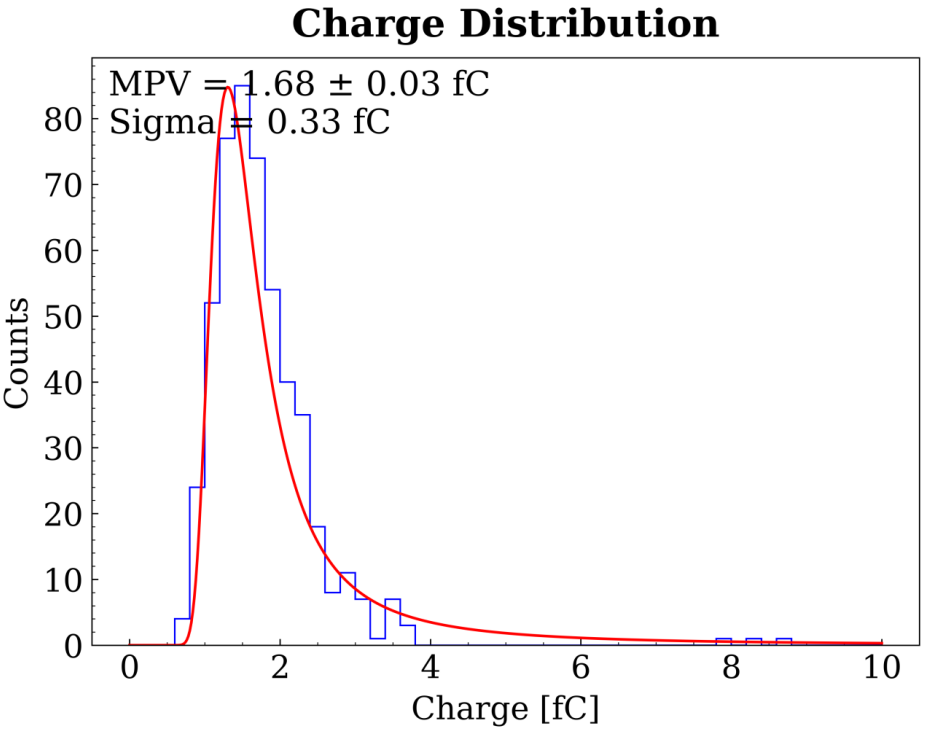}
            \put(-1,71){\textbf{(b)}}
        \end{overpic}
    \end{minipage}%
    \hspace{1cm}%
    \begin{minipage}[t]{0.40\textwidth}
        \centering
        \begin{overpic}[width=\textwidth]{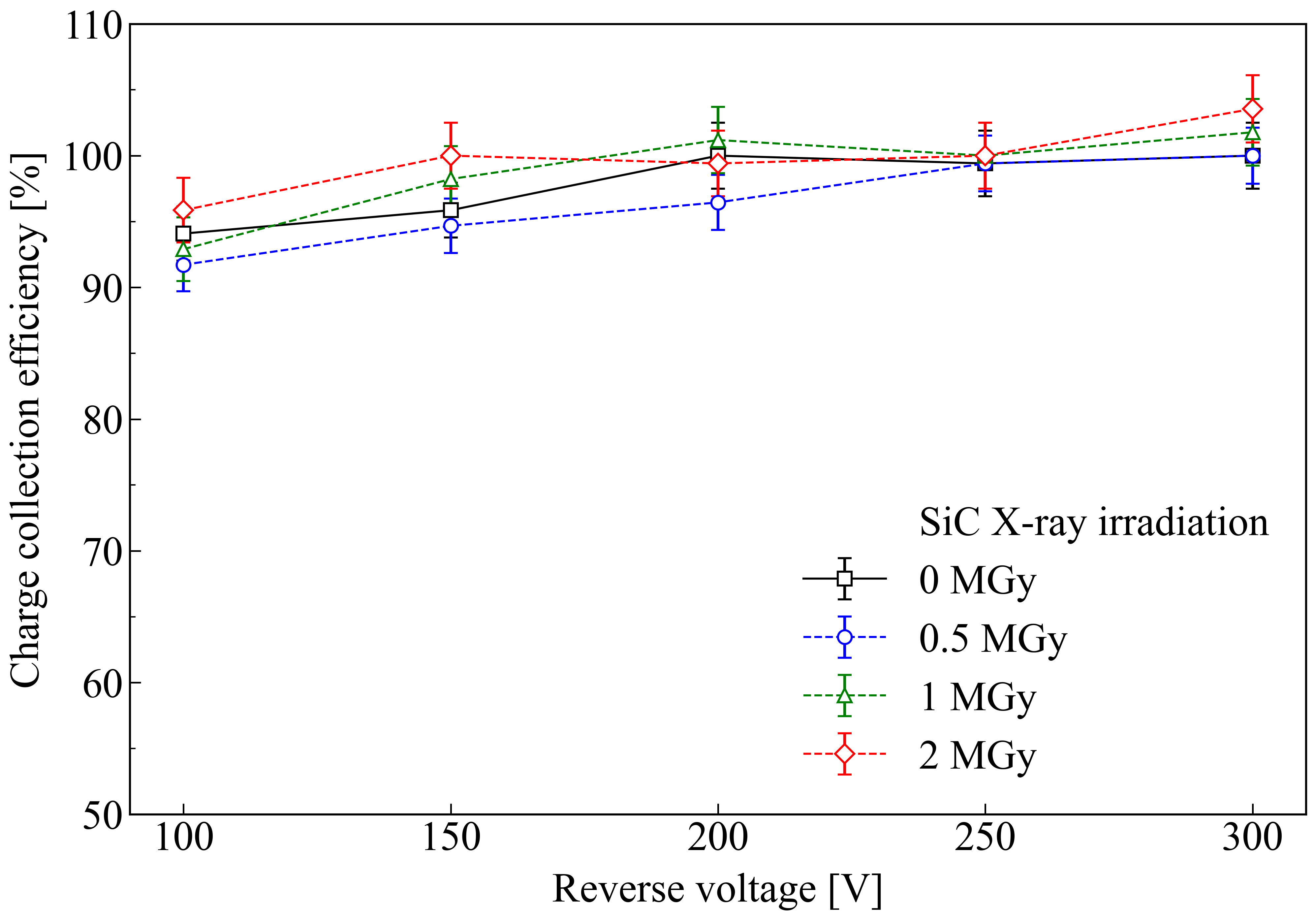}
            \put(-3,65){\textbf{(c)}}
        \end{overpic}
    \end{minipage}
    
    \caption{Charge collection performance of 4H-SiC PIN detectors: (a) Experimental setup; (b) Landau fit of the collected charge spectrum, with MPV indicating characteristic charge; (c) CCE versus reverse bias for different irradiation doses.}
    \label{fig3}
\end{figure*}

To evaluate the charge collection performance of 4H-SiC PIN detectors under X-ray irradiation, measurements were conducted using $\beta$-particles from a \(^{90}\text{Sr}\) source before and after irradiation. The experimental setup, illustrated in Fig.~\ref{fig3}(a), consisted of the radioactive source, the device under test, a single-channel readout board with integrated transimpedance amplifier, a Keithley 2470 high-voltage bias supply, a GWINSTEK GPD-3303S low-voltage power module, and a Tektronix MSO64 oscilloscope.

During the measurements, the oscilloscope acquired the output waveforms with a bandwidth of \SI{2.5}{GHz}. The collected charge was obtained by integrating the voltage-time curve and dividing the result by the transimpedance gain \(R_f\) of the readout board. The baseline offset was determined from the pre-signal region and subtracted from the entire waveform to ensure accurate integration.

The $\beta$-particle signals, which follow a Landau distribution due to their energy loss straggling, were fitted with a Landau function as shown in Fig.~\ref{fig3}(b). The most probable value (MPV) of each fit was taken as the characteristic collected charge under the corresponding condition. This approach accounts for the asymmetric nature of energy deposition in thin detectors, where the Landau distribution exhibits a long tail toward higher energies.

\FloatBarrier

The charge collection efficiency (CCE) was studied as a function of reverse bias voltage for devices irradiated to different total doses: 0, 0.5, 1, and \SI{2}{MGy}. The results are presented in Fig.~\ref{fig3}(c). The CCE values are normalized to the collected charge of the unirradiated device at \SI{300}{V} bias, which is defined as 100\%. For the pristine device, the CCE rapidly increases with bias voltage and saturates above the full depletion voltage of approximately \SI{130}{V}, reaching near 100\% collection efficiency. This behavior is consistent with complete depletion of the active region, where the entire \SI{50}{\mu m} thickness contributes to signal generation.

After irradiation, the CCE exhibits only a slight reduction. Even at the highest dose of \SI{2}{MGy}, the CCE at \SI{300}{V} remains above 95\% of its pre-irradiation value. The minimal degradation across all bias points demonstrates that the 4H-SiC PIN detector maintains excellent charge-collection performance for beta particles even under extremely high X-ray doses. This radiation-induced signal loss is significantly lower than the typical degradation observed in silicon-based detectors under comparable conditions, where CCE can decrease by 50\% or more due to carrier trapping at radiation-induced defects. The superior performance of SiC highlights the substantial advantage of this wide-bandgap semiconductor for applications in harsh radiation environments.

\subsection{Timing Resolution}
\label{sec:timing}

In addition to charge collection, the timing resolution of the detector is a critical parameter for applications requiring precise event timing, such as time-of-flight measurements in medical imaging or nuclear facility monitoring. The timing resolution represents the uncertainty in measuring the particle passage time and directly impacts the ability to resolve events in high-rate environments.

The timing resolution was measured using a dual-channel coincidence method \cite{ECFA2021}. A reference detector with known timing resolution (silicon LGAD, \SI{34}{ps} at \SI{80}{V} bias \cite{Li2021}) was placed in coincidence with the SiC PIN detector under test. Both detectors were exposed to $\beta$-particles from the \(^{90}\text{Sr}\) source. The signals from both channels were read out using identical UCSC boards followed by PE15A1008 main amplifiers (20 dB gain). A 1-meter cable was added to the SiC PIN channel to provide approximately \SI{5}{ns} delay for trigger discrimination.

For each coincidence event, the timing information was extracted using offline constant fraction discrimination (CFD) to eliminate time walk caused by amplitude variations. The CFD method determines the trigger time at a constant fraction (70\%) of the signal amplitude, making the timing independent of pulse height. The process involves:
\begin{enumerate}
    \item Smoothing the waveform to reduce noise
    \item Finding the maximum amplitude \(V_{\text{max}}\) and its time \(t_{\text{max}}\)
    \item Setting the threshold at \(V_{\text{th}} = 0.7 \times V_{\text{max}}\)
    \item Locating the point where the rising edge first crosses \(V_{\text{th}}\) and interpolating to obtain the precise CFD time
\end{enumerate}

The time difference \(\Delta t = t_{\text{DUT}} - t_{\text{Ref}}\) was recorded for thousands of events. The distribution of these time differences was fitted with a Gaussian function to obtain its standard deviation \(\sigma(\Delta t)\). The timing resolution of the device under test, \(\sigma_{\text{DUT}}\), was then calculated using:
\begin{equation}
\sigma(\Delta t) = \sqrt{\sigma_{\text{DUT}}^2 + \sigma_{\text{Ref}}^2}
\label{eq:timing}
\end{equation}
where \(\sigma_{\text{Ref}} = \SI{34}{ps}\) is the known resolution of the reference LGAD detector.

The total timing resolution of a detector system can be further decomposed into several contributions:

\begin{equation}
\sigma_{\text{total}}^2 = \sigma_{\text{jitter}}^2 + \sigma_{\text{TDC}}^2 + \sigma_{\text{walk}}^2 + \sigma_{\text{others}}^2
\label{eq:timing_decomp}
\end{equation}

where \(\sigma_{\text{jitter}}\) arises from electronic noise, \(\sigma_{\text{TDC}}\) is the quantization error of the time-to-digital converter (negligible due to the high sampling rate of \SI{10}{GS/s}), \(\sigma_{\text{walk}}\) is the time walk due to amplitude variations (largely eliminated by the CFD method), and \(\sigma_{\text{others}}\) includes contributions from signal distortion and Landau fluctuations.

The jitter contribution can be estimated using the well-known formula \cite{ECFA2021}:

\begin{equation}
\sigma_{\text{jitter}} = \frac{N}{dV/dt} \approx \frac{t_{\text{rise}}}{S/N}
\label{eq:jitter}
\end{equation}

where \(N\) is the RMS noise level, \(dV/dt\) is the slew rate of the signal at the threshold, \(t_{\text{rise}}\) is the signal rise time, and \(S/N\) is the signal-to-noise ratio. This equation shows that jitter is inversely proportional to the signal slope and directly proportional to noise—higher signal amplitudes and faster rise times lead to lower jitter.

The pristine device showed an excellent timing resolution of \SI{21}{ps} at \SI{300}{V} bias. This value is significantly better than recently reported timing performance of 35–61 ps for 4H-SiC LGADs \cite{Yang2025}, demonstrating the superior intrinsic timing capability of our PIN structure. 

After irradiation to \SI{2}{MGy}, the timing resolution remained remarkably good at \SI{31}{ps}, as shown in Fig.~\ref{fig:timing_trend}(b). Over the same dose range, the jitter increased moderately from \SI{9.9}{ps} to \SI{11.1}{ps} (see Fig.~\ref{fig:timing_trend}(a)), primarily due to the reduced signal-to-noise ratio. The rise time distribution remained stable, indicating that the carrier transport properties were not significantly degraded by irradiation.

These results confirm that the 4H-SiC PIN detector not only maintains its charge collection capabilities but also preserves its fast-timing performance under extreme radiation exposure.

\begin{figure*}[!t]
    \centering
    \begin{minipage}[b]{0.8\textwidth}
        \centering
        \begin{overpic}[width=\textwidth]{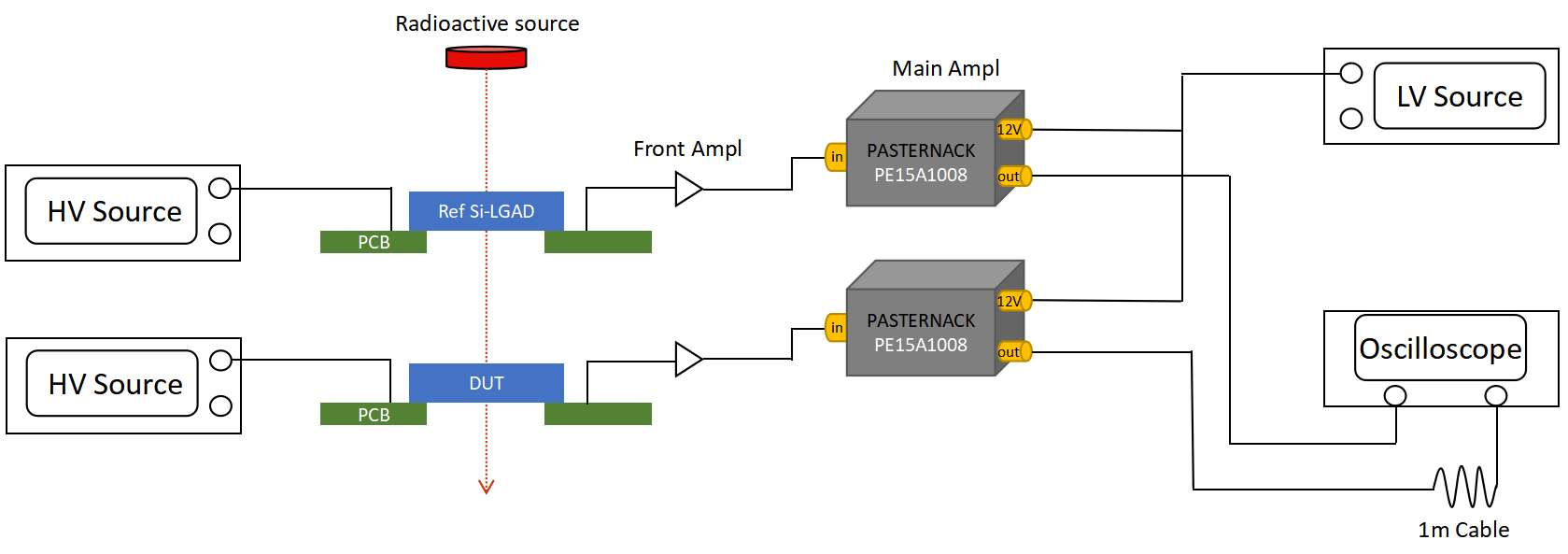}
            \put(-3,32){\textbf{(a)}}
        \end{overpic}
    \end{minipage}\\[1ex]
    
    \begin{minipage}[t]{0.40\textwidth}
        \centering
        \begin{overpic}[width=\textwidth]{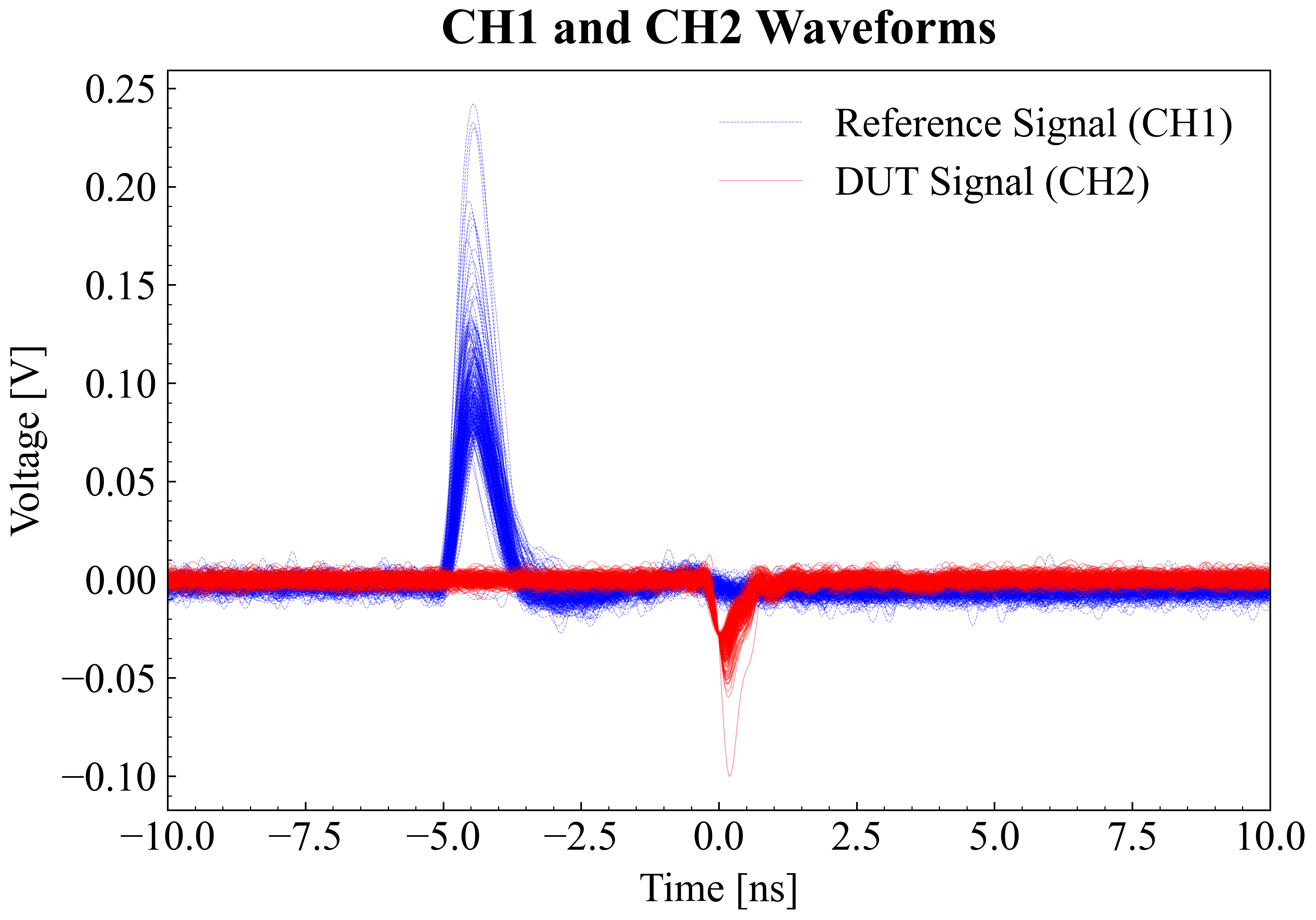}
            \put(-2,62){\textbf{(b)}}
        \end{overpic}
    \end{minipage}%
    \hspace{1cm}%
    \begin{minipage}[t]{0.36\textwidth}
        \centering
        \begin{overpic}[width=\textwidth]{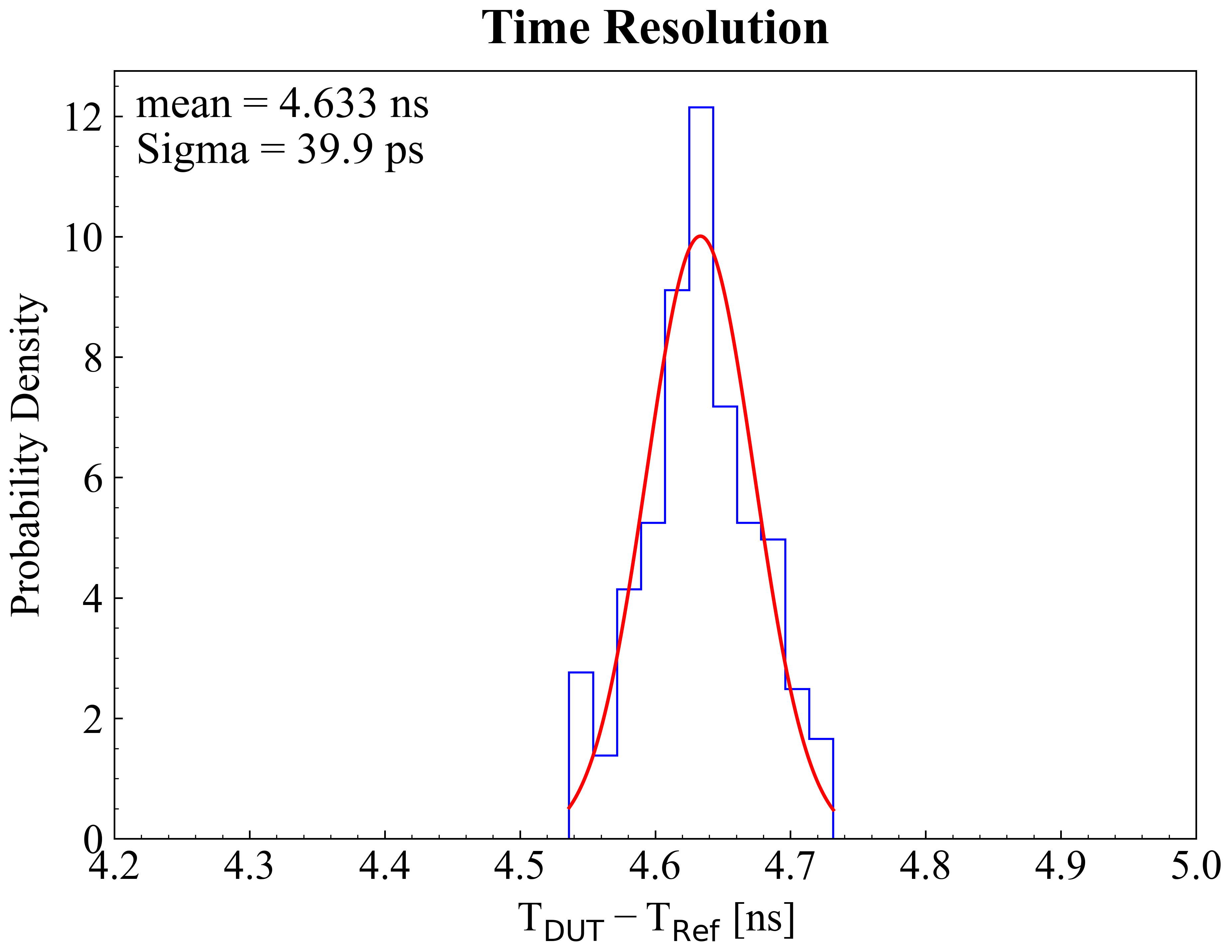}
            \put(-2,69){\textbf{(c)}}
        \end{overpic}
    \end{minipage}
    
    \caption{Timing performance of the unirradiated 4H-SiC PIN detector at \SI{300}{V} bias: (a) Experimental setup; (b) Signal waveforms; (c) Time difference distribution fitted with Gaussian.}
    \label{fig:timing_details}
\end{figure*}

\subsection{Discussion}
\begin{figure*}[!t]
    \centering
    \begin{minipage}[t]{0.4\textwidth}
        \centering
        \begin{overpic}[width=\textwidth]{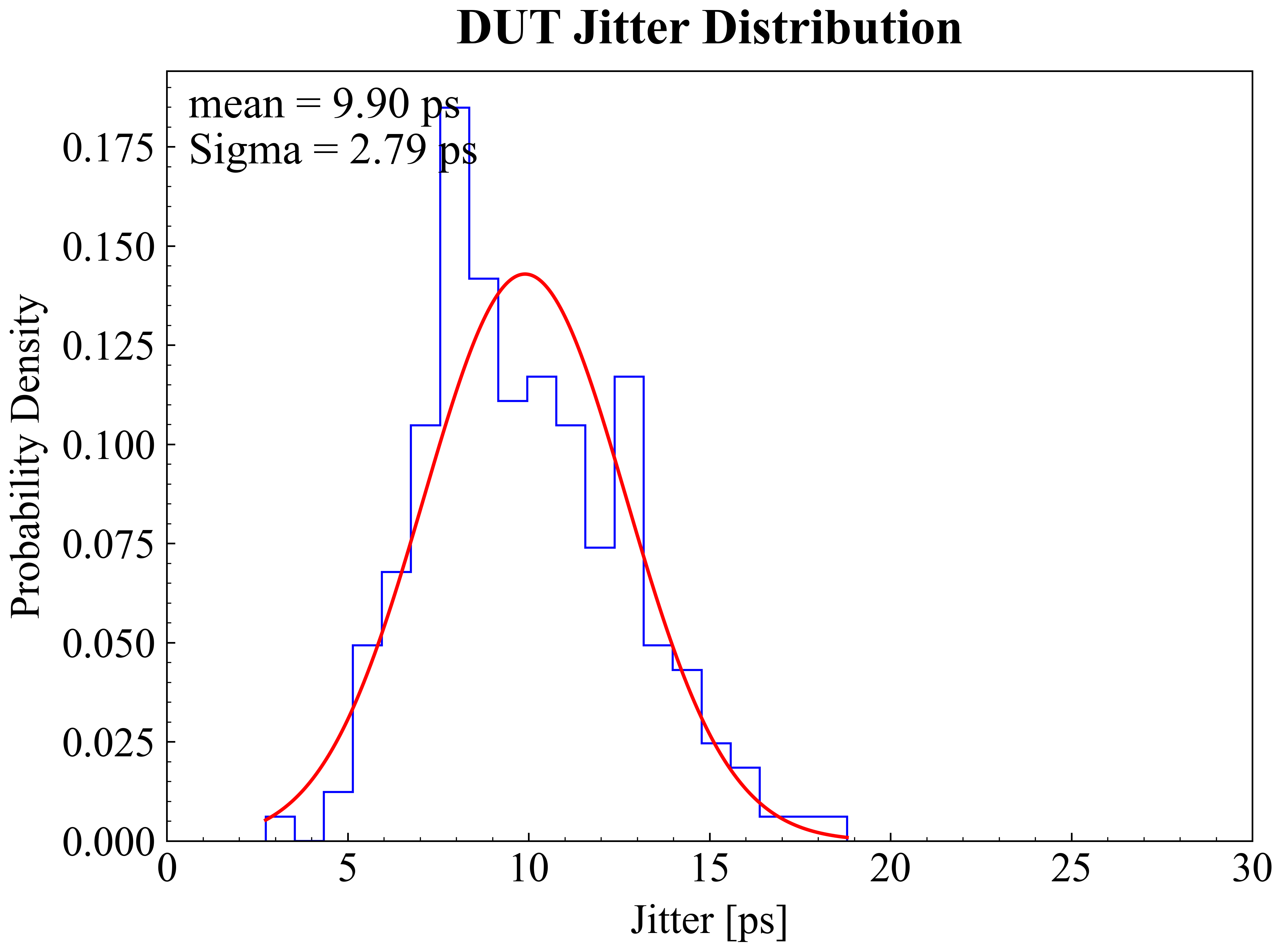}
            \put(-2,69){\textbf{(a)}}  
        \end{overpic}
    \end{minipage}%
    \hspace{1cm}%
    \begin{minipage}[t]{0.4\textwidth}
        \centering
        \begin{overpic}[width=\textwidth]{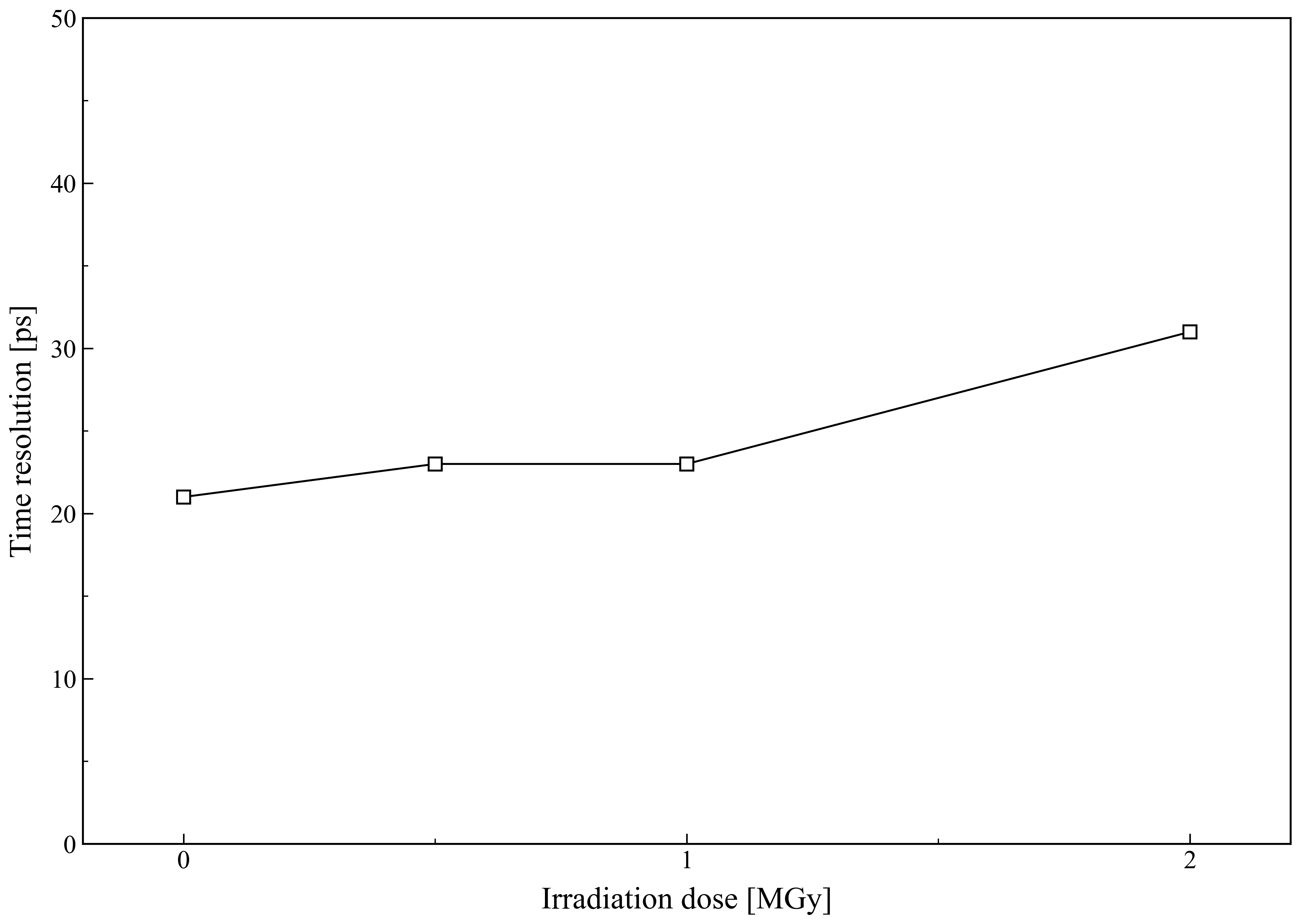}
            \put(-3,70){\textbf{(b)}}  
        \end{overpic}
    \end{minipage}

    \caption{(a) Jitter distribution extracted from noise analysis; (b) Timing resolution as a function of X‑ray irradiation dose.}
    \label{fig:timing_trend}
\end{figure*}

\noindent As shown in Fig.~\ref{fig:timing_trend}(b), the timing resolution degrades only slightly up to \SI{2}{MGy}, remaining well below \SI{50}{ps}. This minimal degradation, together with the stable leakage current and high CCE, confirms the exceptional radiation tolerance of 4H-SiC PIN detectors.

The experimental results consistently demonstrate the exceptional radiation tolerance of 4H-SiC PIN detectors under high-dose X-ray irradiation. The minimal degradation in leakage current, stable C-V characteristics, high charge collection efficiency (\textgreater95\% after \SI{2}{MGy}), and preserved timing resolution (\SI{31}{ps}) collectively indicate that 4H-SiC is a highly promising material for radiation-hard detector applications.

The superior performance of SiC compared to silicon can be understood from fundamental material properties. The wide bandgap of \SI{3.26}{eV} results in an intrinsic carrier concentration approximately 19 orders of magnitude lower than that of silicon, making thermal generation negligible even after irradiation-induced defect introduction. The high displacement threshold energy (25-35 eV for SiC vs. 13-20 eV for Si) means that higher energy is required to create permanent lattice damage, reducing the defect introduction rate under irradiation. Additionally, the high saturation drift velocity enables fast signal collection even in the presence of radiation-induced trapping centers.

The slight degradation in timing resolution after \SI{2}{MGy} irradiation (from \SI{21}{ps} to \SI{31}{ps}) is primarily attributed to the reduced signal-to-noise ratio rather than fundamental changes in carrier transport. The jitter contribution increased from \SI{9.9}{ps} to \SI{11.1}{ps}, consistent with the observed reduction in signal amplitude. This interpretation is supported by the stable rise time distribution, indicating that carrier mobility and saturation velocity remain largely unaffected by the irradiation dose.

To our knowledge, this is the first systematic report demonstrating that 4H-SiC PIN detectors can maintain sub-50 ps timing resolution after exposure to MGy-level X-ray irradiation. This combination of radiation hardness and fast timing capability positions 4H-SiC as a compelling candidate for next-generation detectors in radiation-hard environments such as nuclear facility monitoring, space missions, and medical imaging.

\FloatBarrier

\section{Conclusion}
\label{sec:conclusion}

This study systematically evaluates the radiation tolerance of 4H-SiC PIN detectors under \SI{160}{keV} X-ray irradiation up to \SI{2}{MGy} (Si equivalent). Results demonstrate that even after the maximum dose, the detectors maintain excellent performance: reverse leakage current remains at $\sim 10^{-11}$ \si{A/cm^2} with no degradation, all samples achieve consistent full depletion at $\sim$\SI{130}{V}, charge collection efficiency for beta particles retains $>95\%$ of its pre-irradiation value, and timing resolution shows only a modest increase from \SI{21}{ps} to \SI{31}{ps}, still exhibiting excellent characteristics. These findings mark the first demonstration of sub-50 ps timing resolution in 4H-SiC PIN detectors after MGy-level X-ray exposure, significantly outperforming conventional silicon detectors under equivalent conditions.

The exceptional radiation hardness demonstrated here has profound implications for critical application domains requiring X-ray detection in extreme environments. In nuclear reactor monitoring, advanced X-ray imaging techniques enable non-destructive examination of irradiated fuel and structural components---recent innovations have demonstrated the ability to perform rapid CT scans of nuclear materials, reducing scan times by up to 90\% while enhancing worker safety by minimizing radiation exposure. These methods allow researchers to detect swelling, cracks, and separation in fuel cladding, ensuring geometric stability and safe reactor operation. In space exploration and astronomy, X-ray detectors are essential for missions studying high-energy cosmic phenomena---from cubesats capturing real-time X-ray polarization information from gamma-ray bursts to observatories like IXPE revealing magnetic field configurations around black holes and neutron stars. Future missions demand detectors capable of simultaneous polarization, spectroscopy, and timing measurements under constant cosmic radiation bombardment. In high-energy physics and synchrotron facilities, next-generation experiments require picosecond timing precision while withstanding extreme radiation doses for time-resolved studies of material dynamics. The results experimentally confirm the inherent radiation hardness of SiC and provide critical data support for detector applications across these demanding environments.

Future work will investigate higher-dose irradiations with neutrons and protons, and explore 4H-SiC LGAD designs with built-in gain to further enhance timing performance while preserving radiation hardness.

\end{document}